\documentclass[aps,twocolumn,showpacs,preprintnumbers,amsmath,amssymb]{revtex4}

\usepackage{graphicx}
\usepackage{dcolumn}
\usepackage{bm}

\begin{document}

\preprint{}

\title{A supersonic beam of cold lithium hydride molecules}

\author{S. K. Tokunaga}
\author{J. O. Stack}
\author{J. J. Hudson}
\author{B. E. Sauer}
\author{E. A. Hinds}
\author{M. R. Tarbutt}
\email{m.tarbutt@imperial.ac.uk}

\affiliation{Centre for Cold Matter, Blackett Laboratory, Imperial
College London, SW7 2BW, United Kingdom}

\date{\today}

\begin{abstract}
We have developed a source of cold LiH molecules for Stark
deceleration and trapping experiments. Lithium metal is ablated
from a solid target into a supersonically expanding carrier gas.
The translational, rotational and vibrational temperatures are
$0.9\pm0.1$\,K, $5.9\pm 0.5$\,K and $468\pm17$\,K respectively.
Although they have not reached thermal equilibrium with the
carrier gas, we estimate that $90\%$ of the LiH molecules are in
the ground state, $X^1 \Sigma^+ (v=0, J=0)$. With a single 7\,ns
ablation pulse, the number of molecules in the ground state is
$4.5 \pm 1.8 \times 10^7$ molecules per steradian. A second,
delayed, ablation pulse produces another LiH beam in a different
part of the same gas pulse, thereby almost doubling the signal. A
long pulse, lasting 150\,$\mu$s, can make the beam up to 15 times
more intense.
\end{abstract}


\maketitle

\section{Introduction}
Techniques are currently being developed to form cold and
ultracold molecules for applications across many fields of
research \cite{Doyle04, Bethlem03}. The applications include tests
of fundamental symmetries and time variation of fundamental
constants, quantum information processing, quantum chemistry and
the physics of dipolar quantum degenerate gases. Cold molecules
have more to offer than cold atoms, because of their vibrational
and rotational structure. Polar molecules are of particular
interest, their interaction with electric fields being typically
2-4 orders of magnitude larger than that of atoms. This
interaction is the basis of an important technique for producing
cold molecules, known as Stark deceleration \cite{Bethlem99}.
Molecules seeded in a gas pulse are cooled by supersonic expansion
to temperatures of order 1\,K, but have a centre-of-mass velocity
of several hundred m/s. This is reduced in the Stark decelerator
by a time-varying inhomogeneous electric field. Once brought to
rest, the molecules can be loaded into a trap \cite{Bethlem00} or
a storage ring \cite{Crompvoets01}. To date, the deceleration
technique has been applied to isotopic species of CO
\cite{Bethlem99}, NH$_{3}$ \cite{Bethlem00}, OH \cite{Bochinski03,
Meerakker05}, NH \cite{Meerakker06}, H$_2$CO \cite{Hudson05} and,
in our laboratory, YbF \cite{Tarbutt04}. The ratio, $R$, of Stark
shift to mass serves as a figure of merit for Stark deceleration.
At 200\,kV/cm, $R=0.189$\,cm$^{-1}$/amu for OH and
0.114\,cm$^{-1}$/amu for ND$_{3}$, the only two species to have
been brought to rest and trapped so far. It is a major challenge
to form beams of cold molecules that have suitably large values of
$R$ at practical electric field strengths. The easiest molecules
to produce at low temperature using a supersonic expansion
technique are those that have substantial vapour pressure at room
temperature. However, many of these are unsuitable for
deceleration because $R$ is too low. Conversely, those species
that are most amenable to deceleration may be difficult to produce
in the first instance. The lithium hydride molecule falls into
this second class. It is an excellent candidate for Stark
deceleration, having $R=0.388$\,cm$^{-1}$/amu at 200\,kV/cm.
However, despite the large volume of literature on LiH, there are
no reports of low temperature beams.

The spectroscopy of lithium hydride has been very extensively
studied, both theoretically and experimentally, over many decades.
LiH is sufficiently simple that highly accurate {\em ab initio}
calculations of its energy level structure and physical properties
are possible. For this reason, and because of its low mass, it is
an excellent system for studying the breakdown of the
Born-Oppenheimer approximation \cite{Vidal82, Bellini95}. LiH is
also of considerable astrophysical interest, particularly for its
role in the evolution of the early universe \cite{Bodo03, Puy98}.
In 1993, Stwalley and Zemke published a comprehensive review of
experimental and theoretical work on the three lowest electronic
states of LiH: X$^{1}\Sigma ^{+}$, A$^{1}\Sigma^{+}$ and
B$^{1}\Pi$ \cite{Stwalley93}. More recent studies include
extensive sub-Doppler laser spectroscopy of the A-X system
\cite{Bouloufa00}, high precision studies of the ground-state by
emission spectroscopy in the far and mid-infrared \cite{Bellini95,
Dulick98}, and the first observation and characterization of the
C$^{1}\Sigma^{+}$ state \cite{Lin97, Chen99}. All the experimental
studies of gas-phase LiH have used hot sources. The molecules were
either studied directly inside a heated cell, or an effusive beam
was formed, usually by passing hydrogen gas through a vapour of
lithium produced inside a heated crucible. A supersonic expansion
source of LiH was also developed, but the rotational temperature
of the molecules produced was approximately 600\,K, far higher
than anticipated \cite{Dagdigian76}. The formation of the molecule
in the gas-phase is difficult because, although the overall
reaction ${\rm Li} + (1/2){\rm H}_{2} \rightarrow {\rm LiH}$ is
exothermic by 0.21\,eV, the single-step reaction ${\rm Li} + {\rm
H}_{2} \rightarrow {\rm LiH} + {\rm H}$ is endothermic by 2.05\,eV
when the reactants are in their ground states
\cite{CRCDissociation}. One is thus faced with the challenge of
producing the molecules at low temperatures while simultaneously
overcoming the 24000\,K barrier to the reaction. In this paper, we
report the successful production of sub-Kelvin LiH beams.

\begin{figure*}
\includegraphics{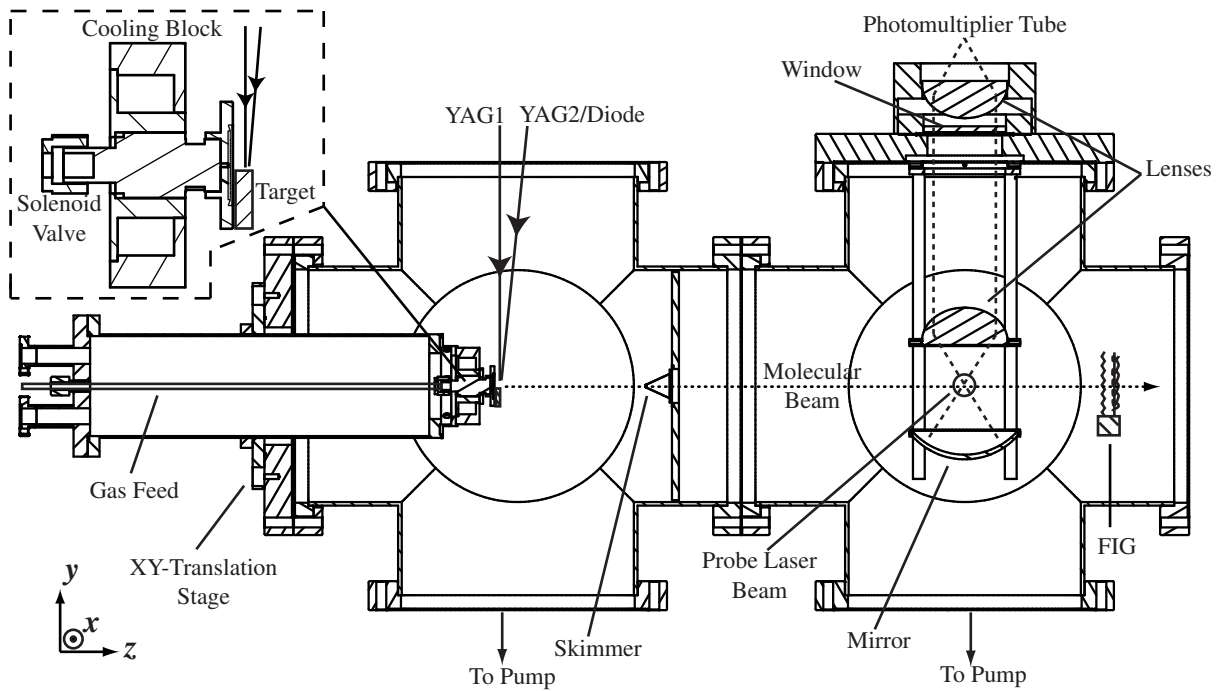}
\caption{\label{Fig:setup} Layout of the experiment and
magnification of the source region (inset). YAG1 is the primary
Nd:YAG laser used for laser ablation of the target. YAG2/Diode
indicates the beam-path of either (i) a second Nd:YAG laser used
for dual-pulse ablation experiments or (ii) a diode laser resonant
with the Li $2s-2p$ transition. The probe laser beam propagates
into the page, along $x$.}\end{figure*}

\section{Experiment Setup}

The experimental setup is shown in Fig. \ref{Fig:setup}. The
vacuum system consists of two DN200CF 6-way crosses, evacuated by
two 1000\,l/s turbomolecular pumps and separated by a skimmer with
a 2\,mm diameter opening. A solenoid valve with a 1\,mm diameter
nozzle (General Valve, Series 99) is inserted into the source
chamber and holds the carrier gas at a pressure in the range
2-6\,bar. A voltage pulse of typical amplitude 250\,V and typical
duration 300\,$\mu$s is applied to the valve, and a pulse of gas
is emitted with an approximately Gaussian profile whose length is
$\sim$100\,$\mu$s \cite{Tarbutt02}. When operated at 10\,Hz
repetition rate, as in all our experiments, the time-averaged
pressures in source and detection chambers are 10$^{-4}$\,mbar and
3$\times$10$^{-7}$\,mbar respectively. The temperature of the
solenoid valve can be controlled with 1\,K precision in the range
100-293\,K, by regulating a flow of nitrogen gas through a cooling
block that surrounds the valve. This assembly is mounted on the
end of a tube, allowing the insertion depth of the valve and hence
the distance from valve to skimmer to be varied. For the
experiments reported below, we fixed this distance at 110\,mm. The
transverse position of the entire valve assembly can be adjusted
under vacuum.

The LiH molecules are produced by laser ablation of a solid Li
target mounted directly onto the face of the valve, and displaced
from the nozzle by approximately 1\,mm along $-y$ (see
Fig.\,\ref{Fig:setup}). The target is ablated using light from a
pulsed Nd:YAG laser (YAG1) providing up to 450\,mJ of energy at
1064\,nm or 110\,mJ at 355\,nm in a 7\,ns pulse (Q-switched mode)
or a train of 1\,$\mu$s pulses lasting for approximately
150\,$\mu$s (long-pulse mode). To investigate dual-pulse
laser-ablation, light from a second 10\,ns, 1064\,nm Nd:YAG laser
(YAG2), is introduced at a small angle to the first. The delays
between valve, YAG1 and YAG2 triggers are computer-controlled with
50\,ns resolution.

The pulsed molecular beam passes through the skimmer into the
detection chamber. Here it encounters the LiH detector 271\,mm
downstream of the skimmer. A fast ionization gauge (FIG) situated
59\,mm further downstream measures the temporal profile of the gas
pulse with a resolution of 5$\mu$s.

In the LiH detector, the molecules are detected by time-resolved
laser-induced fluorescence in the ultra-violet (uv). The laser
system consists of a single-mode cw Ti:Sapphire laser (Coherent
MBR110) pumped by an 8W Nd:YVO$_4$ laser (Coherent Verdi V8) and
frequency-doubled inside a build-up cavity (Coherent MBD200). The
laser frequency is measured to an accuracy of 600\,MHz using a
wavemeter (HighFinesse WS-6). The probe laser beam propagating
along $x$ and linearly polarized along $z$, excites LiH on single
rotational components of the A$^{1}\Sigma^{+}(v') - $X$^{1}\Sigma
^{+}(v''=0)$ transitions. The choice of A-state vibrational
quantum number, $v'$, is based on the efficiency for driving the
transition with the laser power available. With increasing $v'$
(in the range $v'=0-8$), the required laser power decreases due to
the improving Franck-Condon overlaps, but the power available from
the laser system also decreases as the transition frequency moves
deeper into the ultra-violet. In our experiments, we have used
$v'=2,3,4$ at wavelengths of 376.8, 372.0 and 367.2\,nm
respectively. The upper state radiative lifetime is
$\simeq$30\,ns, and non-radiative decay processes are negligible
\cite{Zemke(1)78}. The spectrum of the resulting fluorescence will
contain discrete lines with significant intensity ranging from the
excitation wavelength in the uv out to 1100\,nm. Light-collecting
optics, consisting of a spherical mirror (90\,mm diameter, 30\,mm
focal length) and two condenser lenses (73\,mm diameter, 55\,mm
focal length), form a 1:1 image of this fluorescence at the
8\,mm$\times$24\,mm photocathode of a photomultiplier tube (PMT,
Hamamatsu R928P) operated in photon-counting mode and aligned with
the long axis parallel to $x$. Scattered laser light is suppressed
using a glass filter (Schott GG400) placed in front of the
photocathode. On the $v'=3$ transition, this filter transmits
$<$10$^{-5}$ of the photons at the laser frequency, while
transmitting 73\% of all the fluorescence. The photon rate at the
PMT is measured as a function of the time since firing YAG1, with
up to 1\,$\mu$s resolution. The resulting time-of-flight (TOF)
profile provides a measurement of the molecules' speed and
translational temperature, and the relative intensity of
rotational lines in the spectrum measures the rotational
temperature.

In addition to the probe laser system described above, we also use
a 10\,mW, 670\,nm diode laser (Hitachi HL6714G) tuned to the D1 or
D2 resonance lines of atomic lithium. This laser has two uses. In
the detection region, it serves as a probe laser for lithium,
allowing us to measure the Li flux in the beam. When directed onto
the target, it allows us to investigate the formation of LiH by
the reaction of excited-state Li with H$_{2}$.

\section{Results}

\begin{figure}
\includegraphics{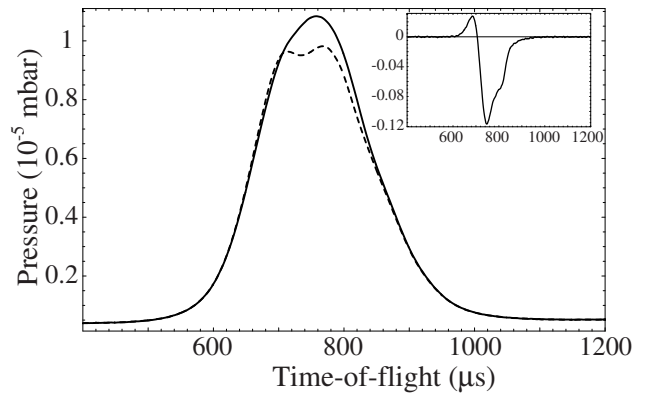}
\caption{\label{Fig:FIG} Gas pulse measured on the fast-ionization
gauge with Nd:YAG turned off (solid line) and on (dashed line).
The inset shows the difference between the two
profiles.}\end{figure}

\subsection{Single-pulse ablation}
We begin by discussing the particular case of single-pulse
1064\,nm laser ablation of a Li target into a carrier gas of
Ar(98\%)/H$_{2}$(2\%) at 3\,bar stagnation pressure and 293\,K
stagnation temperature. The ablation laser is Q-switched,
providing a 7\,ns pulse. Figure \ref{Fig:FIG} shows the typical
signal observed on the fast-ionization gauge under these
conditions. To produce this data, we modulated between YAG1-off
(solid line) and YAG1-on (dashed line). The data shows that the
ablation process results in a local `hole' in the gas pulse. The
depth and width of this hole increases with increasing YAG1 power,
and the position of the hole relative to the centre of the gas
pulse is fixed by the time delay between valve and YAG1 triggers.
For the data shown, this time delay was chosen to be 485\,$\mu s$
in order to place the hole near the centre of the gas pulse. It is
known that, following ablation, the plume forms and propagates to
the gas jet on a sub-microsecond timescale \cite{Labazan06}. We
therefore take the trigger for YAG1 as marking the moment when the
hole is formed. This is the origin of time in Fig.\,\ref{Fig:FIG}.
The arrival time of the gas pulse at the FIG determines its
central speed as 582\,m/s. As in other experiments using this type
of valve, this speed is a little higher than the expected terminal
speed for an ideal supersonic expansion of the gas, which is
557\,m/s. The inset in Fig.\,\ref{Fig:FIG} is the difference
between the signal with the ablation laser on and off. It shows a
small increase in the gas density at short times, followed by a
much larger decrease at longer times. Collisions between the hot,
energetic ablation plume and the gas pulse result in a loss of
atoms from the carrier gas pulse, causing the local decrease in
the density. A natural interpretation of the increase at short
times is that additional energy deposited through interactions
with the ablation plume leads to a slightly higher terminal
velocity. Qualitatively, this type of behaviour is observed for a
wide range of ablation power, valve voltage, valve-to-YAG timing
and YAG-target alignment, though the detailed shape of the profile
changes considerably.

\begin{figure}
\includegraphics{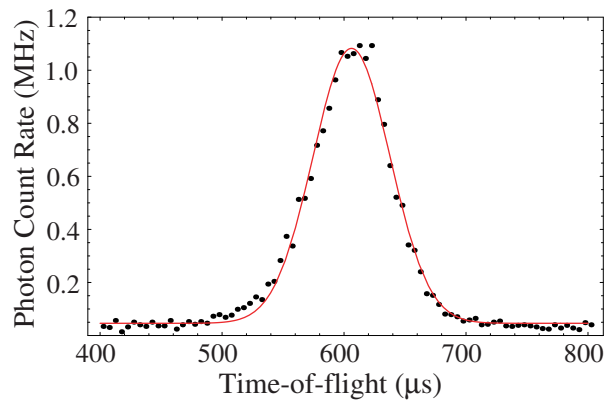}
\caption{\label{Fig:TOF} LiH time-of-flight profile and gaussian
fit.}\end{figure}

Figure \ref{Fig:TOF} shows the time-of-flight (TOF) profile
observed at the LiH detector with the probe laser tuned to excite
the $^{7}$Li$^{1}$H A$^{1}\Sigma^{+}(v'=3) - $X$^{1}\Sigma
^{+}(v''=0)$ R(0) transition at 805798\,GHz. The signal is the
average over 100 shots from the source, representing 10 seconds of
integration time. The abscissa records the time since firing the
ablation laser, while the ordinate gives the photon rate measured
at the PMT. We fit a Gaussian (solid line) to the time-of-flight
data, this being the expected profile when the translational
temperature is low and the distribution formed at the source is
narrow \cite{Tarbutt02}. The molecules reach their terminal speed
in a time much smaller than their total time-of-flight to the
detector, and so the centre of the distribution immediately gives
the mean speed. Since, as we show below, the pulse of molecules
has a very narrow initial width ($\sim5\mu$s), the width of the
profile at the detector is an accurate measure of the
translational temperature. The speed and translation temperature
obtained from Fig.\,\ref{Fig:TOF} are 628\,m/s and 1.0\,K
respectively. We observe slow drifts in the speeds and
temperatures obtained by this method which we attribute primarily
to changes in target condition. Analyzing several datasets taken
under nominally identical conditions many hours apart, we measure
a mean speed of 625$\pm$5\,m/s, and a mean translational
temperature of 0.9$\pm$0.1\,K. The velocity slip between the LiH
and the carrier gas is a common feature of seeded molecular beams.

\begin{figure}
\includegraphics{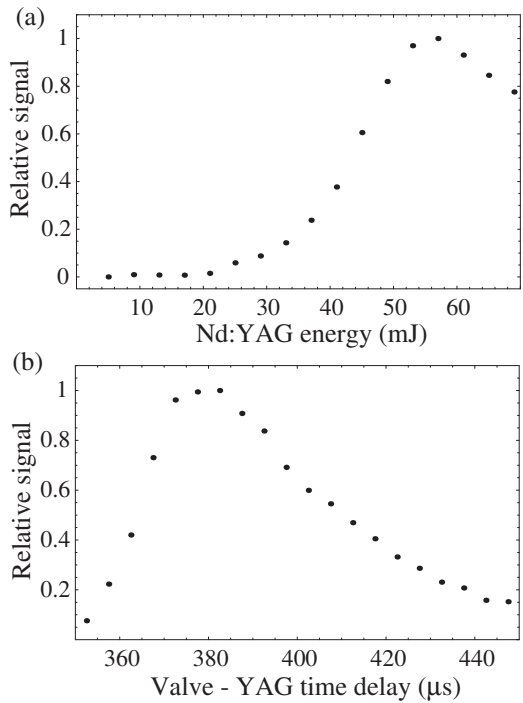}
\caption{\label{Fig:Scalings} Scaling of the ground-state LiH beam
intensity with (a) Nd:YAG pulse energy and (b) time delay between
valve trigger and Nd:YAG Q-switch trigger.}
\end{figure}

Figure \ref{Fig:Scalings}(a) shows how the LiH signal varies with
Nd:YAG pulse energy. The signal peaks at an ablation energy near
55\,mJ per pulse, coinciding with the appearance of the `hole' in
the gas pulse roughly as shown in Fig. \ref{Fig:FIG}. This
observation suggests that the signal-limiting process involves a
trade-off between the ablation yield and the preservation of the
gas-pulse. At low energies the ablation yield, and hence the LiH
signal, is small, while at high energies the ablation plume
destroys the gas pulse, again leading to a small LiH signal. It
follows that a pulsed valve delivering higher peak gas densities
in correspondingly shorter pulses might yield higher LiH beam
intensities.

There is a time delay between the valve trigger and the appearance
of the gas pulse above the target due to the mechanical response
of the solenoid valve. To optimize the LiH signal, the time delay
between valve and YAG1 triggers needs to be chosen correctly.
Figure \ref{Fig:Scalings}(b), a plot of our signal as a function
of this time delay, shows that the optimal delay is 385\,$\mu$s.
We note that this optimum is approximately 100\,$\mu$s shorter
than the delay used to centre the hole in the gas pulse, as in
Fig. \ref{Fig:FIG}. This means that it is optimal to form the
ablation plume when the leading edge of the gas pulse is above the
target. Still shorter time delays result in higher translational
temperatures as well as lower intensity.

\begin{figure*}
\includegraphics[width=13cm]{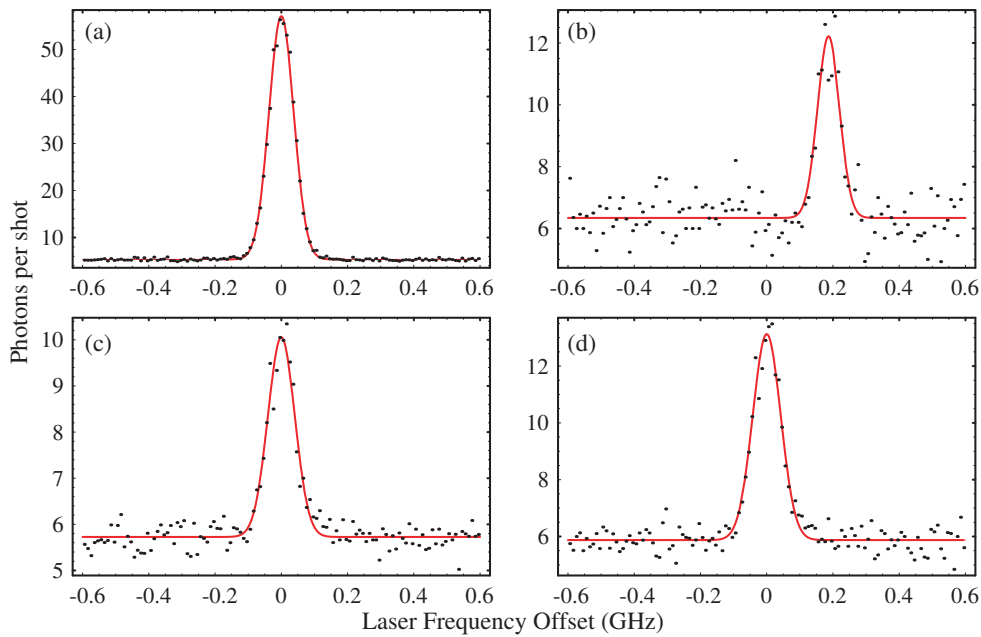}
\caption{\label{Fig:Spectra} Probe laser scans around the
resonances A$^{1}\Sigma^{+}(v') - $X$^{1}\Sigma ^{+}(v'')$ R($J$),
along with Gaussian fits. (a) $^{7}$LiH, $v'=3, v''=0, J=0$, (b)
$^{6}$LiH, $v'=3, v''=0, J=0$, (c) $^{7}$LiH, $v'=3, v''=0, J=1$
and (d) $^{7}$LiH, $v'=6, v''=1, J=0$. The central frequencies of
these resonances, as recorded by a wavemeter with $\pm$600\,MHz
absolute accuracy, are (a) 805798.1\,GHz, (b) 805966.4\,GHz, (c)
805702.4\,GHz and (d) 797837.6\,GHz. These values are all in
agreement with those tabulated in \cite{Bouloufa00}, or, where not
tabulated, calculated from the set of mass-dependent Dunham
coefficients given in the same reference.}\end{figure*}

By integrating under the TOF profile in Fig. \ref{Fig:TOF} we
measure the total number of photons detected per shot with the
probe laser tuned to resonance. Figure \ref{Fig:Spectra}(a) shows
the data obtained upon scanning the laser frequency several times
over a 1.2\,GHz region around this resonance. The Gaussian fit to
this data (solid line) gives a FWHM of 88\,MHz, due to the Doppler
width arising from the divergence of the detected beam. We observe
very similar resonances when the laser is tuned through the R(0)
components of the corresponding $v'=2$ and $v'=4$ transitions at
795734.0\,GHz and 816375.0\,GHz respectively \cite{Bouloufa00}.
Figure \ref{Fig:Spectra}(b) shows the same transition as in (a),
but for the $^{6}$LiH isotopic species. Taking into account the
change of laser intensity between datasets (a) and (b), we
determine the ratio of $^{6}$Li to $^{7}$Li in our source to be
7.7$\pm$0.5\%, consistent with natural Li abundances. In order to
determine the ground-state fraction in our beam, and hence the
rotational temperature, we scanned over the $^{7}$LiH R(1)
resonance, as shown in Fig. \ref{Fig:Spectra}(c). The sum of the
squared matrix elements for this transition is twice that of the
R(0) line. Taking this, and the change of laser intensity into
account, we find the population ratio between the ground and first
rotationally excited states to be $N(J=1)/N(J=0)=8.2 \pm 2.5 \%$.
Assuming thermal equilibrium amongst the rotational states, the
corresponding rotational temperature is $T_{\rm rot} = 5.9 \pm
0.5$\,K, significantly hotter than the translational temperature.
Furthermore, by analyzing the time-of-flight profiles, we observe
that the $J=1$ molecules have a speed that is $5.9 \pm 1.3$\,m/s
higher, and a translational temperature that is $0.25 \pm 0.07$\,K
higher than the $J=0$ molecules in the same beam. This correlation
indicates that the translational degree of freedom has not reached
thermal equilibrium with the carrier gas and that the rotationally
excited molecules are further from thermal equilibrium than those
in the ground state. A search for the R(2) spectral line at
805336\,GHz did not yield a signal, allowing us to set an upper
bound on the population ratio $N(J=2)/N(J=0)<4 \%$. This is
consistent with the rotational temperature determined above, but
does not help to constrain its value further.

Turning now to the vibrations of the molecule, it is well known
that this degree of freedom often decouples from the expansion at
a much higher temperature than the others. To investigate this, we
scanned the probe laser over the $^{7}$LiH A$^{1}\Sigma^{+}(v'=6)
- $X$^{1}\Sigma ^{+}(v''=1)$ R(0) transition, and obtained the
data shown in Fig. \ref{Fig:Spectra}(d). Knowing the laser
intensities and the Franck-Condon factors \cite{Zemke(2)78}, we
deduce the population ratio between the first vibrationally
excited state and the ground state to be $N(v''=1)/N(v''=0)=1.5
\pm 0.2 \%$. The corresponding vibrational temperature (assuming
thermal equilibrium between the vibrations) is $468 \pm 17$\,K. To
investigate population in higher-lying vibrational levels, we
scanned a 4\,GHz region centred on the $^{7}$LiH
A$^{1}\Sigma^{+}(v'=10) - $X$^{1}\Sigma ^{+}(v''=2)$ R(0)
transition at 804948\,GHz, but did not observe a signal. The
corresponding upper bound on the $v''=2$ population is
$N(v''=2)/N(v''=0)<1 \%$, consistent with, but not further
constraining, the above vibrational temperature.

Finally in this section, we mention that ablation at 355\,nm,
rather than at 1064\,nm, does not change significantly the LiH
signal obtained. We also note that we have not fully understood
the LiH formation mechanism, particularly the role played by the
H$_2$ in the carrier gas. We find that decreasing the H$_2$
concentration does not result in a decreasing LiH signal. Indeed,
the signal remains upon switching the carrier gas to pure argon,
even after thoroughly flushing the gas-handling system. This
puzzling observation suggests that it is hydrogen released from
the target that is primarily responsible for LiH formation, rather
than hydrogen present in the carrier gas.

\subsection{Beam intensity}

We now estimate the ground state LiH flux in our beam, using the
signal measured on the $v'=3$ R(0) resonance. To ensure uniform
laser power within the interaction volume, we expanded and
collimated the probe beam and then selected the central portion
using an aperture of height (along $y$) $H=4$\,mm and width (along
$z$) $W=3$\,mm. The power passing through this aperture was
$P=2.3$\,mW.

The number of photons detected per shot is given by
\begin{equation}
p=\frac{1}{L^2}\int\int N(x,y) \epsilon(x,y)
s(x,y)\,dx\,dy,\nonumber
\end{equation}
where $N(x,y)$, $\epsilon(x,y)$ and $s(x,y)$ are the number of
ground state molecules per steradian, the detection efficiency and
the number of photons scattered by each ground state molecule, all
in the interval $dx\,dy$ around the point $(x,y)$ in the plane of
detection. $L$ is the distance from source to detector. The values
of $N$, $\epsilon$ and $s$ are independent of $y$ over the small
interval defined by the probe laser, $H$. At the laser intensity
used, there is little power broadening, and only those molecules
whose transverse Doppler shifts are within the natural linewidth
contribute significantly to the signal. These molecules are
confined to a small region about the beam axis within which $N$
and $\epsilon$ are independent of $x$. With these simplifications,
we have
\begin{equation}
p=N\epsilon (H/L^2)(L/v_z) \int s(v_x) dv_x,\nonumber
\end{equation}
where we have used the substitution $x=L v_x/v_z$, $v_x$ and $v_z$
being the speeds in the $x$ and $z$ directions.

$s$ is a strongly varying function of $v_x$ due to the transverse
Doppler shift. The upper state lifetime is much shorter than the
time spent in the detector, and the probability of an excited
molecule returning to the ground vibrational state is only 3\%. It
follows that the number of photons scattered by a molecule with
transverse speed $v_x$ is well approximated by
$s(v_x)=1-\exp[-R(v_x) \tau]$, $R(v_x)$ being the excitation rate
for this molecule, and $\tau$ the time taken to traverse the probe
beam. When the laser is on resonance,
\begin{equation}
R\,\tau=\frac{D^{2}}{\epsilon_{0}c\,\hbar^{2}}\,\frac{\gamma}{\Delta^{2}+\gamma^{2}}
\frac{P}{H v_z}.\nonumber
\end{equation}
Here, $\Delta = 2\pi v_{x}/\lambda$ is the Doppler shift, $2\gamma
= 32.4\times10^6$\,s$^{-1}$ is the natural linewidth of the
transition \cite{Zemke(1)78}, and $D^2=|\langle A,v'=3,J'=1|e
z|X,v''=0,J''=0 \rangle|^2$ is the squared matrix element of the
dipole operator between the initial and final states. The latter
can be factored into a rotational part whose value is $1/3$ and an
electronic-vibrational part whose value is $0.0486(e a_{0})^{2}$
\cite{Zemke(2)78}. Evaluating the integral, we obtain
$p=6.6\times10^{-5}N\epsilon$.

The detection efficiency may be written as
\begin{equation}
\epsilon=(\Omega_{\rm l} /4\pi) \sum_i q_i (1+{\cal
R}(\lambda_i))T_{\rm l}(\lambda_i)^2 T_{\rm w}(\lambda_i)T_{\rm
f}(\lambda_i)\chi(\lambda_i).\nonumber
\end{equation}
Here, $q_i$ is the fraction of fluorescent photons in the emission
line whose wavelength is $\lambda_i$. ${\cal R}$, $T_{\rm l}$,
$T_{\rm w}$, $T_{\rm f}$ and $\chi$ are respectively the
wavelength-dependent mirror reflectivity, lens transmission,
window transmission, filter transmission and PMT quantum
efficiency. $\Omega_{\rm l}$ is the solid angle subtended by the
light-gathering lens (1.27 steradians). The result is $\epsilon =
0.96\%$ giving $p=6.4\times10^{-7}N$. We measured $p=29$ photons
per shot, giving us $N=4.5 \times 10^7$ molecules per steradian
per shot. We estimate the relative uncertainty on this value to be
$\pm35\%$, taking into account the uncertainties in all the
quantities involved in the calculation, including variations we
observe from one target spot to another (which is the dominant
contribution to the uncertainty).

The signal gradually decays as the ablated spot on the target
ages, reaching half its initial value after about 20,000 shots.

\subsection{Resonant excitation of the Li $2p$ state}
As mentioned above, the reaction ${\rm Li} + {\rm H}_{2}
\rightarrow {\rm LiH} + {\rm H}$ is endothermic by 2.05\,eV when
the reactants are in the ground state. This value is reduced to
$0.20$\,eV when the lithium is excited to the $2p$ state. It
therefore seemed possible that excitation of the Li would result
in an enhancement of the LiH yield. When investigating this
reaction in a cell heated to approximately 800\,K, Myers {\it et
al.} \cite{Myers87} did indeed find a large enhancement of the
reaction rate when the Li was resonantly excited on the D$_1$ and
D$_2$ transitions at 670.8\,nm; they measured the corresponding
reaction cross-section to be $0.10 \pm 0.03 {\rm \AA}^{2}$. In our
experiments, we scanned the frequency of a 10\,mW diode laser over
the D$_1$ and D$_2$ resonances and looked for an increase in the
LiH yield. No such increase was observed, suggesting either that
the reactants are sufficiently energetic for the reaction to
proceed efficiently in the ground state, or that this reaction
channel is not the main LiH production mechanism in the source.

\subsection{Dual-pulse ablation}

\begin{figure}
\includegraphics{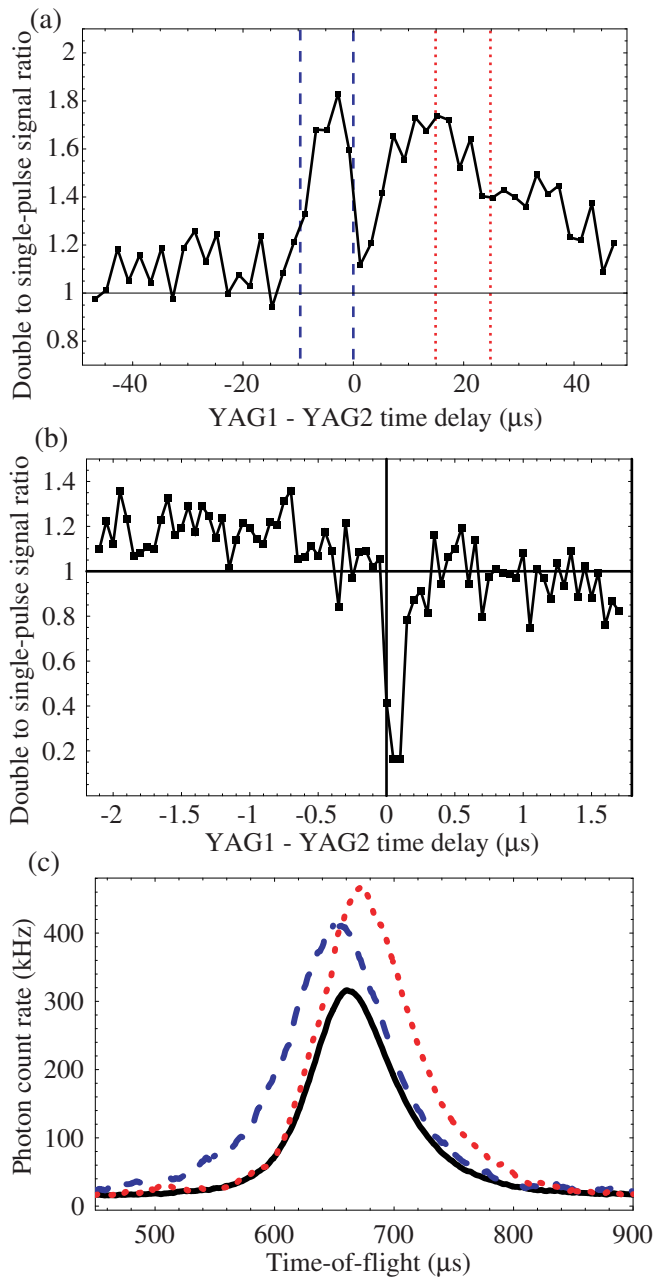}
\caption{\label{Fig:Dual} Dual-pulse ablation results. (a)
Double-pulse to single-pulse signal ratio as a function of the
time delay between the two ablation pulses scanned with 2\,$\mu$s
resolution.  (b) As (a), but with 50\,ns resolution in the
interval around zero time-delay. (c) Time-of-flight profiles for
single-pulse ablation (solid line), dual-pulse ablation with
time-delays between -10\,$\mu$s and 0\,$\mu$s (dashed line) and
with time-delays between 15\,$\mu$s and 25\,$\mu$s (dotted line).}
\end{figure}

Recent literature on laser-induced breakdown spectroscopy (LIBS)
of solid targets reports large signal enhancements when a
dual-pulse ablation scheme is used (e.g.\,\cite{Hohreitera05} and
references therein). Typically, the two pulses are temporally
separated by a few $\mu$s. Enhancements are reported for co-linear
geometries where both laser beams hit the target, and in
orthogonal geometries where one beam travels over the surface of
the target, which the other hits. The underlying mechanisms for
these enhancements are the focus of current research in this area.
Since our own experimental setup shares some features with the
LIBS technique, we searched for a similar signal enhancement using
the (near) co-linear dual-pulse arrangement shown in Fig.
\ref{Fig:setup}.

We scanned the time-delay between the firing of YAG1 and YAG2,
modulating YAG2 on and off so as to obtain the ratio of dual-pulse
to single-pulse LiH beam intensity at each time delay. The
single-pulse signal was first optimized with respect to YAG1
energy, YAG2 energy, position on the target and time-delay between
the valve and YAG1 triggers. Figure \ref{Fig:Dual} shows typical
results. The data were taken with energies and spot sizes of
42\,mJ, 4\,mm (YAG1) and 19.5\,mJ, 2\,mm (YAG2). Figure
\ref{Fig:Dual}(a) shows the signal ratio as a function of the time
delay scanned with 2\,$\mu$s resolution. Over most of the interval
from -10 to +40\,$\mu$s, there is a clear increase in the signal
with the addition of YAG2. We believe this is due to the formation
of a second LiH cloud in a different part of the gas pulse: each
plume interacts with only a narrow ($\sim5\,\mu$s) slice of the
gas. This idea is supported by the 5\,$\mu$s wide region near zero
time delay where there is no increase in the signal. Here, the
second plume has little effect because the gas is already
optimally depleted by the first, as discussed in the context of
Fig.\,\ref{Fig:Scalings}(a). The asymmetry between negative and
positive time-delays in Fig.\,\ref{Fig:Dual}(a) arises because the
pulse from YAG1 interacted with the early part of the gas pulse,
rather than the central part (as discussed above, this choice
optimizes the single-pulse intensity).

We re-scanned the region around zero time delay with 50\,ns
resolution. This revealed the much narrower dip shown in Figure
\ref{Fig:Dual}(b), when the first and second YAG pulses coincide
within 250\,ns. The signal decreases to about 20\% of the
single-pulse value because the effective ablation intensity is
approximately twice the optimum value when the two pulses hit the
target almost simultaneously. We have observed much broader
signal-drops at higher values of YAG2 energy.

For each data point in Figs.\,\ref{Fig:Dual}(a,b) there is a
corresponding time-of flight profile. In \ref{Fig:Dual}(c), the
solid line shows the TOF profile measured for single-pulse
ablation. The dashed line shows the dual-pulse TOF profile
averaged over the data points between -10\,$\mu$s and 0\,$\mu$s
(the region between the dashed lines in (a)), while the dotted
line is the corresponding average for the interval 15\,$\mu$s to
25\,$\mu$s (indicated by dotted lines in (a)). We see that when
YAG2 fires before YAG1 the profile is broadened on the
early-arrival side, and when it fires later it is broadened on the
late-arrival side. The molecules leave the source as two separate
pulses which, due to their finite temperature, merge into a single
broadened pulse as they travel down the beamline. When YAG2 is
delayed by more than 50\,$\mu$s, we observe two distinct peaks in
the TOF.

\subsection{Long-pulse ablation}

The logical extension of dual-pulse ablation is multi-pulse, or
long-pulse ablation, with a pulse length chosen to match the
duration of the gas pulse. In this way, the entire gas pulse would
contribute to the molecular signal. One convenient way to achieve
this is to operate the Nd:YAG laser without a Q-switch. In this
case, the laser undergoes relaxation oscillations emitting a pulse
train that lasts for approximately 150\,$\mu$s and consists of
approximately 100 pulses, each of $\sim1\,\mu$s duration. The
total energy output is roughly the same as in Q-switched mode.

In order to observe a LiH signal in this mode, we found it
necessary to compensate for the much lower ablation power by
focussing the laser onto the target using a lens placed one focal
length (500\,mm) away from the target. We then observed a
ground-state LiH signal whose intensity increased with increasing
ablation energy up to the maximum energy available. The maximum
size of the signal obtained was approximately 15 times larger than
in Q-switched mode. Because of the much broader temporal
distribution of molecules at the source, we are not at present
able to comment on the translational temperature. The greatly
increased signal is accompanied by very large fluctuations on the
timescale of tens of seconds. We attribute these fluctuations to
the drastic changes in surface conditions that occur as the
tightly focussed ablation laser forms deep holes and broad craters
in the target. Initial experiments indicate that these large
fluctuations can be minimized by continuously translating the
target. We are currently modifying our source so as to take full
advantage of the much larger signals obtainable in this long-pulse
mode.

\subsection{Conclusions}
We have produced a beam of LiH with a translational temperature
below 1\,K. We observe a number of non-equilibrium effects in our
beam: velocity slip, a rotational temperature higher than the
translational temperature, an even higher vibrational temperature,
and a rotational-state dependence on the speed and translational
temperature. Despite these effects, the majority of the molecules
are cooled to the ground state. In a first step towards increased
beam intensity, we find that long-pulse ablation produces a far
higher flux than single-pulse ablation, due to a more uniform
loading of the carrier gas pulse. We are exploring several
promising avenues to increase the intensity further - various
targets and hydrogen donor gases, ablation inside an extended
nozzle, and electrical discharge to dissociate the molecular
hydrogen. We plan to use our cold LiH beam in deceleration and
trapping experiments, with a view to reaching sub-mK temperatures
using a second-stage cooling scheme such as sympathetic cooling
\cite{Lara06} or cavity-assisted cooling \cite{Domokos02}.

\acknowledgements We are indebted to Jon Dyne for his expert
technical assistance, and to Gerard Meijer, Irena Labazan and Tim
Steimle for very valuable discussions. We acknowledge UK support
from the Royal Society, EPSRC and PPARC, and the Cold Molecules
Network of the European Commission.

\end{document}